\documentclass[aps,pra,twocolumn,superscriptaddress,amsmath,showpacs,10pt]{revtex4-1}
\usepackage{graphicx}
\usepackage{amssymb}
\usepackage{dsfont}
\usepackage{bm}
\usepackage{siunitx}
\usepackage{braket}
\usepackage{hyperref}

\begin{document}
\title{Coincidence landscapes for polarized bosons}
\author{Jizhou Wu}
\email{wujizhou@mail.ustc.edu.cn}
\affiliation{%
	Shanghai Branch, National Laboratory for Physical Sciences at Microscale,
	University of Science and Technology of China, Shanghai 201315, China%
    }

\author{Hubert de Guise}
\email{hubert.deguise@lakeheadu.ca}
\affiliation{Department of Physics, Lakehead University, Thunder Bay, Ontario P7B 5E1, Canada}
\author{Barry C.\ Sanders}
\email{sandersb@ucalgary.ca}
\affiliation{%
	Shanghai Branch, National Research Center for Physical Sciences at Microscale,
	University of Science and Technology of China, Shanghai 201315, China%
    }
\affiliation{%
    Institute for Quantum Science and Technology, University of Calgary, Alberta T2N 1N4, Canada%
    }
\affiliation{%
    Program in Quantum Information Science, Canadian Institute for Advanced Research,
    Toronto, Ontario M5G 1M1, Canada%
    }
\date{\today}
\begin{abstract}
Passive optical interferometry with single photons injected into some input ports and vacuum into others
is enriched by admitting polarization,
thereby replacing the scalar electromagnetic description by a vector theory,
with the recent triad phase being a celebrated example of this richness.
On the other hand,
incorporating polarization into interferometry is known to be equivalent to scalar theory if the number of channels is doubled.
We show that passive multiphoton $m$-channel interferometry described by SU($m$) transformations
is replaced by SU($2m$) interferometry if polarization is included and thus that the multiphoton coincidence landscape,
whose domain corresponds to various relative delays between photon arrival times,
is fully explained by the now-standard approach of using immanants to compute coincidence sampling probabilities.
Consequently, we show that the triad phase is manifested simply as SU(6) interferometry with three input photons,
with one photon in each of three different input ports.
Our analysis incorporates passive polarization multichannel interferometry into the existing scalar-field approach to computing multiphoton coincidence probabilities in interferometry and demystifies the triad phase.
\end{abstract}
\maketitle
\section{Introduction} 
%
\label{sec:introduction}
Two-photon interference has been studied extensively since the celebrated 1987 Hong-Ou-Mandel (HOM) experiment~\cite{Hong1987PRL}.
Also known as the HOM dip,
this phenomenon is manifested as two identical photons whose
relative arrival times are controlled by a time-delay mechanism,
thereby controlling their relative distinguishability.
The photons arrive at a balanced (50:50) beam splitter 
with each entering a different input port.
Light from the two output ports is directed to two photon counters and the resultant signal is multiplied.
In the quantum description,
the HOM dip corresponds to a null coincidence for zero relative time delay:
the quantum mechanical description forbids each detector to see a photon when they are indistinguishable.
For long delays relative to the duration of the photonic wave packet,
each photon has a 50\% chance of being reflected or transmitted so the coincidence probably is one-half.
The HOM phenomenon provides a vital tool for characterizing single-photon sources~\cite{Santori2002Nature} and for preparing and detecting entangled photonic qubits~\cite{Knill2001Nature, Marcikic2003Nature, Su2017PRL}.

This destructive interference between two photons can be generalized to the case of $n$ photons~\cite{Campos2000PRA,Mahrlein2015OE,Lim2005NJP}.
For three-photon interference, each photon is injected into different input ports, 
and the interest is in the coincidence of three-photon interference with one photon in separate output ports.
The coincidence rate appears as a landscape tuned by two variables,
with the time delays of two photons relative to a third, used as reference.
Not only do we obtain the destructive dip,
but we also obtain a constructive peak in the three-photon landscape~\cite{Tan2013PRL,deGuise2014PRA,Tillmann2015PRX}.

A systematic generalization from the two-photon to high-order coincidence dips or peaks 
arises from treating such scattering transformations as  elements of a unitary group.
In this scheme, the lossless beam splitter is characterized as a U(2) transformation [SU(2) with a global phase] for the photon's states~\cite{Yurke1986PRA,Campos1989PRA}. Using the four-port beam splitter as building blocks, an arbitrary U($m$) interferometer with~$m$ input ports
and~$m$ output ports is achieved as a sequence of beam splitters and phase shifters~\cite{deGuise2018PRA,Clements2016Optica,Reck1994PRL,Rowe1999JMP}.
Below we omit the global phase as it has no effect
on the coincidence rate, 
in which case we write SU($m$) rather than U($m$).

In this paper, 
we consider the U($m$) matrix to be fixed and focus on the parameters that come from the injected bosons, 
i.e., the time delays or the polarization spectrum.  
The analysis of coincidence landscapes in our work is thus different from the landscapes considered in~\cite{Mahrlein2015OE},
where these authors have adjustable parameters for the scattering interferometer and their injected photons are indistinguishable.

There are several equivalent approaches to computing the coincidence rates when there are $n$ photons entering an otherwise arbitrary SU($m$) interferometer.
Some approaches rely heavily on permanents~\cite{Shchesnovich2015PRA,Tichy2015PRA,Tamma2015PRL};
they have the advantage of familiarity and a well-known algorithm,
namely, Ryser's algorithm~\cite{Ryser1963},
to evaluate the permanent of any matrix.

An alternative is to recognize that additional (even considerable) insight into many-body problems is often obtained by the use of symmetric-group methods.
Using Schur-Weyl duality between the symmetric group S$_{n}$ of $n$ objects
and the unitary group U($m$) yields an immanant-based formalism that makes coincidence landscapes amenable to interpretations based on permutational symmetries of photons described by 
S$_{n}$~\cite{Tan2013PRL,deGuise2014PRA,Tillmann2015PRX,Khalid2018PRA}.
Immanants naturally account for the partial distinguishability of photons,
and the idea of ``immanon''~\cite{Tichy2017PRA} has been introduced in an attempt to capture the feature.
Although the use of Schur-Weyl duality brings elegance to this alternative approach,
immanants are not so well known and good algorithms to compute them are not common,
even if a close connection (going back to Kostan~\cite{Kostan1995JAMS}) 
between immanants and group functions can be exploited~\cite{Burgisser2000SIAMJC, deGuise2016JPA}.  

Despite this paucity of algorithms, the problem of reducing a representation of~$S_n$ is well studied and can be transformed to an eigenvalue problem using class operators~\cite{Chen2002}.  Most important, this reduction is independent of the details of the scattering network, so that the transformation effecting the block diagonalization can be precomputed with prior knowledge only of the input and output configurations of the system~\cite{Khalid2018PRA}.  An example of
this feature can be found in Eq.~(\ref{eq:Vblockdiagonal}) of Appendix~\ref{app:equivalence}.

Here we generalize previous results~\cite{Tan2013PRL,deGuise2014PRA,Tillmann2015PRX,Khalid2018PRA} to accommodate interference of polarized photons.
The case of more than one photon in an input or output port means that
possible representations of~S$_n$ are a subset of all the possible irreducible representations (irreps)
and is treated in~\cite{Khalid2018PRA}. 
We show in particular that the tools based on the symmetric group 
and developed for the analysis of photons which are temporally partially distinguishable
can be extended to include partial distinguishability in polarization;
in other words,
all the insights provided by the use of immanants remain,
provided we consider both polarization and temporal degrees of freedom (labeled by~$\bm{\theta}$ and $\bm{\tau}$, respectively)
in the distinguishability features of the photons.

We give, in Eq.~(\ref{eq:same_Tichy}) of Appendix~\ref{app:equivalence}, an expression for the rate \begin{equation}
	C(\bm{\theta};\bm{\tau}),
\end{equation}
which agrees with the result of~\cite{Tichy2015PRA},
but can be usefully converted to 
an expression
\begin{equation}
	C(\bm{\theta};\bm{\tau})=\sum_{i,j}u_{i}^*R_{ij}(\bm{\theta};\bm{\tau})u_j,
\end{equation}
where further block diagonalization of $R_{ij}$ will produce a rate 
(using the standard tools of the symmetric group)
in terms of moduli squared of sums of immanants written as combinations of $u_i$'s.

Typically multiphoton interference is treated as a scalar field theory:
polarization is ignored.
Ignoring polarization is not detrimental for studying multiphoton interferometry as polarization can rather trivially be converted to a path degree of freedom.
Polarization is included in $m$-channel interferometry by combining a U(2) transformation with a U($m$) transformation,
resulting in U($2m$) irreps using the subchain
\begin{equation}
	\text{U}(2m)\supset \text{U}(m)\times \text{U}(2)
\end{equation}
transformation and the dual pair of U(2) and U($m$) when the same U(2) polarization transformation acts on all spatial modes~\cite{Rowe1999JMP}.
Dhand and Goyal decomposed arbitrary unitary transformations for photonic states in terms of spatial  and internal modes~\cite{Dhand2015PRA}.
Manipulating internal and external degrees of freedom  in multiphoton interference has recently attracted  significant attention~\cite{Menssen2017PRL,Agne2017PRL}. 

Despite a mathematical equivalence between polarization and path degrees of freedom,
some confusion has arisen about the sufficiency of mutual photon distinguishability in explaining the features of coincidence rates.
As a specific example,
Menssen {\it et al}.\ used a symmetric three-port interferometer, known as a tritter~\cite{Jex1995OC,Zukowski1997PRA},
to implement two- and three-photon interference with polarized photons~\cite{Menssen2017PRL}.
They then claim that ``the distinguishability between pairs of photons is not sufficient to fully describe the photons' behavior in a scattering process,
but that a collective phase, the triad phase, plays a role''~\cite{Menssen2017PRL}.
Agne {\it et al}.~\cite{Agne2017PRL} showed this collective phase in entangled-photon experiments.
A chief goal of our work is to show that
polarization degrees of freedom, independent for each physical mode, are
trivially absorbed into an SU($2m$) transformation.

In Sec.~\ref{sec:from_su},
we show how to construct the special unitary network with doubling of ports when polarization is taken into consideration.
Then we calculate the coincidence rate for polarization-sensitive and -insensitive events in our framework in Sec.~\ref{sec:coincidence_rate}.
In Sec.~\ref{sec:triad_phase_Demystified}, we focus on the case when $m=3$ and clarify the triad phase in our framework.

\section{From SU(\texorpdfstring{$m$}{m}) to SU(\texorpdfstring{$2m$}{2m})} 
\label{sec:from_su}
In this section,
we give the general framework to deal with the polarized multiphoton interference problem in a polarization-sensitive scattering network. We generalize a discussion in~\cite{Rowe1999JMP} to easily incorporate polarization into scalar-field interferometry by converting polarization to path.
If the polarization transformations on each spatial modes are identical,
``one simply extends the U($n$) group to U($n$)$\times$U(2) and to include combinations of polarizers and beam splitters,
for example,
one extends to $\text{U}(2n)\supset \text{U}(n)\times\text{U}(2)$~\cite{Rowe1999JMP}.
''~\cite{Rowe1999JMP}.
Here we consider the polarization transformations being distinct on all the spatial modes.

First,
notice that we can construct a U(2) transformation matrix for the polarization.
For a photon injecting in the $i$th port of an~$m$-channel polarization-sensitive interferometer,
the input state is
\begin{equation}
	\Ket{\Psi_i}
 =\sum_{\alpha=1}^2c^{(i)}_{\alpha}A^\dagger_{i\alpha}(\tau_i)\Ket{0},
\end{equation}
and
\begin{equation}
	A^\dagger_{i\alpha}(\tau_i):=\int \text{d} \omega \phi_i(\omega)\text{e}^{-\text{i} \tau_i\omega} a_{i\alpha}^\dagger(\omega)\, .
\end{equation}
Here $\tau_i$ is the time delay of the $i$th photon;
$a^\dagger_{i\alpha}$ is the creation operator for the photon mode of the $i$th spatial port and $\alpha$th polarization.
Thus, $\alpha=1$ and $\alpha=2$ represent two orthogonal polarizations,
e.g., horizontal and vertical polarizations or left-handed and right-handed polarizations.
\begin{figure}
\centering
\includegraphics[width=\columnwidth]{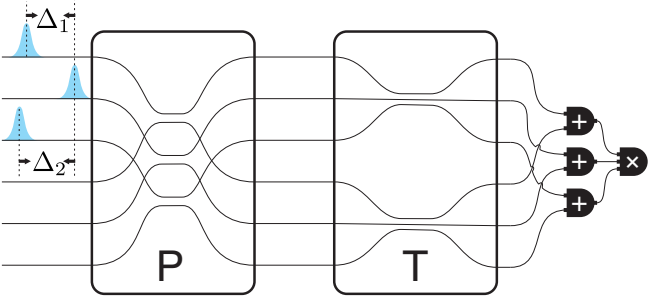}
\caption{Decomposition of interference network and setup for the detectors.}
\label{fig:setup}
\end{figure}

Without loss of generality,
we use $\alpha=1$ to represent horizontal polarization and $\alpha=2$
to represent vertical polarization;
$\vert\phi(\omega)\vert^2$ is the frequency profile in the frequency domain satisfying
\begin{equation}
	\int\text{d} \omega\vert\phi(\omega)\vert^2=1.
\end{equation}
Coefficients $\{c_\alpha^{(i)}\in \mathbb{C}\}$ satisfy
\begin{equation}
	\sum_{\alpha=1}^2\vert c_\alpha^{(i)}\vert^2
		=1.
\end{equation}
These coefficients $\{c_\alpha^{(i)}\}$
describe a U(2) transformation~$P^{(i)}$ on the $i$th photon's polarization as
\begin{equation}
\label{eq:polarization}
P^{(i)}:=
\begin{pmatrix}
 c_1^{(i)} & -\text{e}^{\text{i}\psi}c_2^{*(i)}\\
 c_2^{(i)}& \text{e}^{\text{i}\psi}c_1^{*(i)}
\end{pmatrix},
\end{equation}
where $\psi$ is an overall phase.

For~$m$ spatial modes,
the polarization transformation is
\begin{equation}
\label{eq:ptransformation}
	P:=\oplus_{i=1}^{m}P^{(i)}.
\end{equation}
We model the scattering network $T$ to be polarization sensitive by factoring the net effect of the network into individual polarization transformation for each spatial mode followed by a transformation mixing modes for the same polarization,
which takes the form
\begin{equation}
  \label{eq:Ttransformation}
T:=\tilde{T}^{(1)}\oplus\tilde{T}^{(2)},
\end{equation}
where $\tilde{T}^{(\alpha)}$ is the mode transformation acting on photons with polarization $\alpha$ alone. Here we assume $\tilde{T}^{(1)}=\tilde{T}^{(2)}$ in the respective polarization subspaces.

For convenience,
we specify the basis order by using one single subscript to replace the spatial and polarization subscripts:
$a_k^\dagger
		:=a_{i\alpha}^\dagger$,
with
\begin{equation}
	a^\dagger_k
		=a^\dagger_{i1},\;
	k=i=1,\dots,m
\end{equation}
and
\begin{equation}
	a^\dagger_k=a^\dagger_{i2},\;
	k=i+m, i=1,\dots,m.
\end{equation}
In this subscript notation,
the first $k$ labels up to $k=m$ refer to horizontally polarized photons,
while the next $k$ labels from $m+1$ to $2m$ refer to vertically polarized photons.
For example,
when $m=3,n=2$,
$a^\dagger_{12}$ is replaced by $a^\dagger_{4}$,
which is the creation operator of horizontally polarized photon in the first spatial mode.

Thus,
by ignoring the phase shift in Eq.~(\ref{eq:polarization}) as it will not affect the result of coincidence,
our original SU($m$) scattering network spanned by~$m$ space-mode states is expanded to an $\text{SU}(2m)$ transformation
\begin{equation}
\label{eq:UTP}
	U=TP
\end{equation}
spanned by $2m$ orthogonal space-and-polarization-mode states (see Fig.~\ref{fig:setup} as an example when $m=3$) so that photons scatter according to
\begin{equation}\label{eq:creation_transformation}
a_j^\dagger\mapsto Ua_j^{\dagger} U^{\dagger}=\sum_{i=1}^{2m}a_i^\dagger U_{ij},
\end{equation}
where $a^\dagger_i$ is the creation operator for the photon in the $i$th mode.

Notice that the first index of~$U$ represents output and the second index represents input: a photon in input mode $j$ is scattered to a linear combination of photons existing in output mode~$i$.
The transformation of the creation operators is verified from the photon state's transformation,
\begin{align}
U\ket{1_j}
\label{eq:state_1}=&U a_j^{\dagger}\ket{0}=U a_j^\dagger U^\dagger U\ket{0}=U a_j^\dagger U^\dagger\ket{0}\\
\label{eq:state_2}=&\sum_{i}\ket{1_i}\bra{1_i}U\ket{1_j}=\sum_{i}a_{i}^{\dagger}U_{ij}\ket{0},
\end{align}
where~$\ket{1_i}$ is the state that one photon is in the $i$th mode.
Comparing the last equations in Eqs.~(\ref{eq:state_1}) and~(\ref{eq:state_2}),
we obtain the creation-operator transformation~(\ref{eq:creation_transformation}).
Using Fig.~\ref{fig:setup},
we confirm the factorization $U=T P$ of Eq.~(\ref{eq:UTP}).

For $m=3$ modes, the explicit matrix representation of the SU(6) transformation $T$ can be
\begin{equation}
T=\begin{pmatrix}
\tilde{T}&0\\
0&\tilde{T}
\end{pmatrix}
\end{equation}
with~$\tilde{T}$ the original 3$\times$3 transformation matrix.
By writing the independent
polarization transformations $P^{(1)},P^{(2)},P^{(3)}$ 
as per Eq.~(\ref{eq:polarization}),
we construct the polarization transformation
\begin{align}
P&=P^{(1)}\oplus P^{(2)}\oplus P^{(3)}\\
&=\begin{pmatrix}
P^{(1)}_{11}&0&0&P^{(1)}_{12}&0&0\\
0 &P^{(2)}_{11} & 0 & 0 & P^{(2)}_{12} & 0 \\
0 & 0 & P^{(3)}_{11}  & 0 & 0 & P^{(3)}_{12} \\
P^{(1)}_{21}&0&0&P^{(1)}_{22}&0&0\\
0 & P^{(2)}_{21} & 0 & 0 & P^{(2)}_{22} & 0 \\
0 & 0 & P^{(3)}_{21} & 0 & 0 & P^{(3)}_{22} \\
\end{pmatrix}. \label{eq:explicit_P}
\end{align}
Thus, the total SU(6) transformation is
\begin{widetext}
\begin{equation}
  \label{eq:explicit_U}
	U(\bm{\theta})=T P(\bm{\theta})\\ 
	=
		\begin{pmatrix}
P^{(1)}_{11} \tilde{T}_{11} & P^{(2)}_{11} \tilde{T}_{12} & P^{(3)}_{11} \tilde{T}_{13} & P^{(1)}_{12} \tilde{T}_{11} & P^{(2)}_{12} \tilde{T}_{12} & P^{(3)}_{12} \tilde{T}_{13} \\
P^{(1)}_{11} \tilde{T}_{21} & P^{(2)}_{11} \tilde{T}_{22} & P^{(3)}_{11} \tilde{T}_{23} & P^{(1)}_{12} \tilde{T}_{21} & P^{(2)}_{12} \tilde{T}_{22} & P^{(3)}_{12} \tilde{T}_{23} \\
P^{(1)}_{11} \tilde{T}_{31} & P^{(2)}_{11} \tilde{T}_{32} & P^{(3)}_{11} \tilde{T}_{33} & P^{(1)}_{12} \tilde{T}_{31} & P^{(2)}_{12} \tilde{T}_{32} & P^{(3)}_{12} \tilde{T}_{33} \\
P^{(1)}_{21} \tilde{T}_{11} & P^{(2)}_{21} \tilde{T}_{12} & P^{(3)}_{21} \tilde{T}_{13} & P^{(1)}_{22} \tilde{T}_{11} & P^{(2)}_{22} \tilde{T}_{12} & P^{(3)}_{22} \tilde{T}_{13} \\
P^{(1)}_{21} \tilde{T}_{21} & P^{(2)}_{21} \tilde{T}_{22} & P^{(3)}_{21} \tilde{T}_{23} & P^{(1)}_{22} \tilde{T}_{21} & P^{(2)}_{22} \tilde{T}_{22} & P^{(3)}_{22} \tilde{T}_{23} \\
P^{(1)}_{21} \tilde{T}_{31} & P^{(2)}_{21} \tilde{T}_{32} & P^{(3)}_{21} \tilde{T}_{33} & P^{(1)}_{22} \tilde{T}_{31} & P^{(2)}_{22} \tilde{T}_{32} & P^{(3)}_{22} \tilde{T}_{33}
		\end{pmatrix},
\end{equation}
\end{widetext}
which clearly transforms the SU(3) description into a SU(6) description.

The cost of doubling the dimension of the problem is that the $2m\times 2m$ scattering matrix~$U$ is no longer completely specified by the original $m\times m$ matrix $T$,
but now depends on the parameters of the polarization transformations $P^{(k)}$. 
We denote by~$\bm{\theta}$ the collection of polarization parameters needed to specify the various $P^{(k)}$, 
and hence the argument for $U(\bm{\theta})$ and $P(\bm{\theta})$.  
The benefit of this doubling is that
all the tools available for the discussion of unpolarized photons can simply be imported {\it ipso facto} to the problem that includes polarization.

We conclude this section with a discussion of a special case,
already hinted at earlier.
If the polarization transformations~$P^{(i)}$ of 
Eq.~(\ref{eq:ptransformation}) are all identical, 
then the transformations $T$ and~$P$ 
that define~$U$ in Eq.~(\ref{eq:UTP}) 
actually commute, i.e.,
\begin{equation}
\label{eq:UTPPT}
	U=T P=P T.
\end{equation}
The transformation SU($2m$) then factors into
$\text{SU}(m)\otimes\text{SU}(2)$,
and by combining with the overall phase shift,
we are back to the Rowe {\it et al}.\ case~\cite{Rowe1999JMP}.

This factorization of Eq.~(\ref{eq:UTPPT}) is  of particular interest when photons are fully indistinguishable.
In this case, the representation of SU($2m$) must be ``fully symmetric'',
or transform by an irrep of the type
$(n,0,\dots)$ of SU($2m$).
As explained in~\cite{Rowe1999JMP},
this constrains the possible irreps of $\text{SU}(m)\otimes\text{SU}(2)$:
the SU(2) and SU($m$) representations must be associated with identical Young diagrams~\cite{Hagen1965JMP},
so that by preparing~$n$ photons in a definite state of polarization [i.e., a definite SU(2) irrep]
these photons are simultaneously prepared in SU($m$) irreps  with nontrivial spatial permutation symmetries.
This is the optical analog of combining spin and spatial wave functions of 
specified permutation symmetry to construct a fully symmetric wave function for~$n$ bosons.

\section{Coincidence Rate}
\label{sec:coincidence_rate}
In this section,
we give the coincidence rate for polarized multiphoton interference,
augmenting an earlier discussion where the polarization was not explicitly discussed in the immanant-based formalism~\cite{Tan2013PRL, deGuise2014PRA, Tillmann2015PRX,Khalid2018PRA}. 
Here we give the polarization-insensitive and -sensitive detection cases using the immanant-based formalism.

For polarization-sensitive detection,
we first define the~$n$-dimensional input configuration vector~$\bm{v}$,
whose entry $v_i\in\Set{1,2,\dots,2m}$ indicates which input mode the $i$th photon occupies.
For convenience of description,
we also use a $2m$-dimensional input mode-occupation vector $\bm{\xi}^{\bm{v}}$ to represent the same input state,
whose entry
\begin{equation}
	\xi^{\bm{v}}_i\in\Set{0,1,\dots,n}
\end{equation}
tells how many photons are in the $i$th mode,
and
\begin{equation}
	\sum_{i=1}^{2m}\xi^{\bm{v}}_i=n.
\end{equation}

Similarly,
we define an~$n$-dimensional output configuration vector~$\bm{\eta}$ and a $2m$-dimensional output mode-occupation vector $\bm{\xi}^{\bm{\eta}}$,
with
\begin{equation}
	\eta_i\in\Set{1,2,\dots,2m},\;
	\xi^{\bm\eta}_i\in\Set{0,1,\dots,n},
\end{equation}
and
\begin{equation}
	\sum_{i=1}^{2m}\xi^{\bm{\eta}}_i=n.
\end{equation}
For polarization-insensitive detectors,
we also define another output configuration $\bm{\mu}$, 
called regular output configuration, 
with entry
\begin{equation}
	\mu_i\in\Set{1,2,\dots,m}.
\end{equation}

Now we inject~$n$ photons into the SU($2m$) interferometer~$U$ to obtain the input state
\begin{equation}
\Ket{\Psi_{\text{in}}}:=\mathsf{N}\prod_{i=1}^n A_{v_i}^\dagger(\tau_i)\Ket{0}
\end{equation}
where 
\begin{equation}
\mathsf{N}=\frac{1}{\sqrt{\xi^{\bm v}_1!\,\xi^{\bm v}_2!\,\dots\xi^{\bm v}_{2m}!}}
\end{equation}
is a normalization constant and
\begin{equation}
	A_{v_i}^\dagger(\tau_i):=\int\text{d} \omega\phi_{v_i}(\omega)\text{e}^{-\text{i} \tau_i\omega}a_{v_i}^\dagger(\omega).
\end{equation}
The resultant output state is
\begin{align}
\Ket{\Psi_{\text{out}}}&:=U\Ket{\Psi_{\text{in}}}\\
&=\mathsf{N}\prod_{i=1}^n \int\text{d} \omega\phi_{v_i}(\omega)\text{e}^{-\text{i} \tau_i\omega}\sum_{j=1}^{2m}U_{jv_i}a_j^\dagger(\omega)\Ket{0}.
\end{align}

We next need to specify the measurements for different kinds of detectors.
We divide them into two categories of detectors for polarized single photons:
ideal polarization-sensitive detectors and practical polarization-insensitive detectors.

For the polarization-sensitive detectors,
one polarization-sensitive detector is equivalent to the combination of a polarizing beam splitter and two polarization-insensitive detectors.
Our separation of the polarization from the spatial mode,
which results in the expansion from SU($m$) to SU($2m$),
has played the role of polarizing beam splitters.
Then the detection operator for this specific output configuration~$\bm{\eta}$ is
\begin{equation}
\label{eq:detector_polarization}
	M_{\bm{\eta}}
		:=\frac{1}{\xi^{\bm \eta}_1!\,\xi^{\bm \eta}_2!\,\dots\xi^{\bm \eta}_{2m}!}\int\text{d}
			\bm{\omega}\prod_{i=1}^na_{\eta_i}^\dagger(\omega_i)\Ket{0}\Bra{0}a_{\eta_i}(\omega_i).
\end{equation}

However,
in most boson sampling or multiphoton interference experiments~\cite{Broome2013Science, Spring2013Science, Tillmann2013NatPhotonics, Crespi2013NatPhotonics, Tillmann2015PRX,Wang2017NatPhotonics, Menssen2017PRL},
one usually uses the polarization-insensitive detectors for the single-photon detection,
such as avalanche photodiodes detectors (APDs),
especially single-photon avalanche diode detectors (SPADs).
We deal with the frequency-insensitive detector by summing over the polarizations corresponding to the same regular output configuration $\bm{\mu}$. This set of configurations is
\begin{equation}
	\mathfrak{S}_{\bm\mu}
		:=\Set{\bm{\eta}\mid\forall i\in\Set{1,2,\dots,n},\eta_i\in\Set{\mu_i,\mu_i+m}}
\end{equation}
and
\begin{equation}
	\vert \mathfrak{S}_{\bm\mu}\vert
		=2^n.
\end{equation}
For example,
for a $2\times 2$ spatial scattering interferometer such as a beam splitter,
injecting two photons from input ports 1 and 2,
the input configuration is $\bm{v}=(1,2)$.
Then to detect the coincidence of the photons' output from spatial ports 1 and 2, respectively,
the regular output configuration $\bm{\mu}=(1,2)$,
and
\begin{equation}
	\mathfrak{S}_{\bm{\mu}}=\{(1,2),(1,4),(3,2),(3,4)\},
\end{equation} 
which contains all possible polarizations for the photons' output to spatial ports 1 and 2.
Thus, detection for the regular output configuration $\bm{\mu}$ is
\begin{equation}
\label{eq:measure}
M_{\bm\mu}:=\sum_{\bm{\eta}\in \mathfrak{S}_{\bm\mu}}M_{\bm\eta}.
\end{equation}

We have now formulated how our original problem becomes the problem that~$n$ identically polarized (or horizontally polarized) photons injected into a $2m$-channel interferometer~$U$.  A key point for later discussion is that, in order to proceed, one must \emph{select} an $n\times n$
submatrix $\tilde U(\bm{\theta})$ of $U(\bm{\theta})$ and this submatrix, which is specified by the input and output configurations, and also depends on~$\bm{\theta}$. 

As an example,
consider the $3\times 3$ tritter matrix $T$ given in~\cite{Menssen2017PRL}.
If the input state contains three photons with horizontal polarizations,
and the output state is given by the first two photons with horizontal
and the last with vertical polarization,
the $6\times 6$ matrix $U(\bm{\theta})$ for the polarized network 
collapses to a $3\times 3$ submatrix $\tilde{U}$ given by
\begin{equation}
\tilde{U}(\bm{\theta})=
\frac{1}{\sqrt{3}}
\begin{pmatrix}
P^{(1)}_{11}(\bm{\theta}_1) & P^{(2)}_{11}(\bm{\theta}_2)& P^{(3)}_{11}(\bm{\theta}_3) \\
P^{(1)}_{11}(\bm{\theta}_1) & \zeta^2 P^{(2)}_{11}(\bm{\theta}_2) &  \zeta P^{(3)}_{11}(\bm{\theta}_3)  \\
P^{(1)}_{21}(\bm{\theta}_1)  & \zeta P^{(2)}_{21}(\bm{\theta}_2)  & \zeta^2 P^{(3)}_{21}(\bm{\theta}_3)  \\
\end{pmatrix}
\end{equation}
with $\zeta=\text{e}^{\text{i}\frac{2\pi}{3}}$.
Consequently, experiments where the polarization parameters~$\bm{\theta}$ are changed are equivalent
to experiments where the scattering submatrix $\tilde U(\bm{\theta})$ is changed.
Coincidence rates are therefore expected to depend on~$\bm{\theta}$,
even if the matrix
$\tilde{T}$ that scatters the spatial modes is kept constant from one experiment to the next, and even if the input and output states remain the same as we
vary~$\bm{\theta}$.

For detection modeled by Eq.~(\ref{eq:detector_polarization}),
the coincidence rate for the polarization-sensitive detectors for the output configuration $\bm\eta$ is~\cite{Khalid2018PRA}
\begin{equation}
\label{eq:coincidence_one}
	C_{\bm\eta}(\bm{\theta};\bm\tau):=\bm{u}^\dagger_{\bm\eta}(\bm\theta) R ({\bm\tau})\bm{u}_{\bm\eta}(\bm\theta)
\end{equation}
where $\bm{u}_{\bm{\eta}}(\bm{\theta})$
is an
\begin{equation}
	N_{\bm{\eta}}=\frac{n!}{\xi^{\bm \eta}_1!\,\xi^{\bm \eta}_2!\,\dots\xi^{\bm \eta}_{2m}!}
\end{equation}
dimensional vector,
that contains all the information about the submatrix of our SU($2m$) scattering matrix. 
The vector entries are
\begin{equation}
	\left(\bm{u}_{\bm{\eta}}(\bm{\theta})\right)_k
		=\prod_{i=1}^{n}U(\bm{\theta})_{\bar{\eta}^{(k)}_i v_i}, \text{for}\, k=1,2,\dots,N_{\bm\eta}
\end{equation}
with
\begin{equation}
	\bar{\bm{\eta}}^{(k)}\in\set{\bar{\bm\eta}\vert \bar{\bm\eta}=\sigma{\bm\eta},\,\sigma\in \text{S}_n}.
\end{equation}
The action of $\sigma$ on~$\bm{\eta}$ sends the $i$th element to the $\sigma(i)$th;
for instance,
if $\sigma=(123)$,
then its action on $(1,2,3)$ yields $(3,1,2)$. $\bar{\eta}^{(k)}$s provide a basis of the representation of the symmetric group~S$_n$.

The rate matrix $R(\bm{\tau})$ is
\begin{equation}
	R(\bm{\tau})
    	:=\sum_{\sigma\in S_n}\mathsf{B}_\sigma(\bm{\tau})
        	\Pi_\sigma,
\end{equation}
where~$\Pi_\sigma$ is a permutation matrix with elements
\begin{equation}
(\Pi_\sigma)_{ij}:=\delta(\sigma\bar{\bm{\eta}}^{i},\bar{\bm{\eta}}^{j})=\begin{cases}
1& \sigma\bar{\bm{\eta}}^{i}=\bar{\bm{\eta}}^{j},\\
0 & \text{otherwise}.
  \end{cases}
\end{equation}
For the specific case where there are~$n$ photons in~$n$ distinct output modes, $\Pi_\sigma$ is the $n\times n$ regular representation of $\sigma\in \text{S}_n$. $\mathsf{B}_\sigma$ is the overlap between the input state and the $\sigma$-permuted input state,
\begin{equation}
\label{eq:overlap_term}
	\mathsf{B}_\sigma:=\prod_{i=1}^n\beta_{v_{\sigma(i)},v_i},
\end{equation}
where~$\beta_{i,j}$ is the two-photon pairwise distinguishability between the $i$th and $j$th photons.

Definition~(\ref{eq:overlap_term}) agrees with the distinguishability matrix obtained in~\cite{Tichy2015PRA}. 
Specifically, for the polarization-sensitive case, $\mathsf{B}_\sigma$ 
is only dependent on the time delays $\bm{\tau}$ as $\mathsf{B}_\sigma(\bm{\tau})$ because the overlap is only the spectrotemporal overlap,
\begin{equation}
\label{eq:overlap_1}
\beta_{ij}(\bm{\tau})=\int\text{d} \omega \phi^*_i(\omega)\phi_j(\omega)\text{e}^{\text{i} \omega(\tau_i-\tau_j)}.
\end{equation}

Having given the coincidence rate for the polarization-sensitive detectors in Eq.~(\ref{eq:coincidence_one}),
we now give the coincidence rate for the polarization-insensitive detectors. With the detection operator defined in Eq.~(\ref{eq:measure}), the coincidence is easily obtained as the sum of the coincidence rate of all the configurations in $\mathfrak{S}_{\bm{\mu}}$,
namely,
\begin{equation}
\label{eq:coincidence_original}
	C_{\bm\mu}(\bm{\theta};\bm{\tau})
		:=\sum_{\bm{\eta}\in \mathfrak{S}_{\bm\mu}}C_{\bm\eta}(\bm{\theta};\bm\tau)
=\bm{u}^\dagger_{\bm{\mu}} R (\bm{\theta};{\bm\tau})\bm{u}_{\bm{\mu}}
\end{equation}
The derivation for the last equation is given in Appendix~\ref{app:polarization_coincidence}.

Our key observation is that the effects of the polarization
can be incorporated into a rate matrix
so
\begin{equation}
	R(\bm{\tau})\to R(\bm{\theta};\bm{\tau})
\end{equation}
while making the 
vectors $\bm{u}_\eta(\theta)\to \bm{u}_\mu $ independent of the polarization parameters without affecting the permutation matrices $\Pi_\sigma$.
Specifically, the resulting coincidence rate is the same as~$n$ photons in the original SU($m$) scattering interferometer~$\tilde{T}$,
except that the polarizations are absorbed in the pairwise terms $\beta_{ij}(\bm\tau) \to \beta_{ij}(\bm{\theta};\bm{\tau})$ as
\begin{align}
	\beta_{ij}(\bm{\theta};\bm{\tau})
		=&\sum_{k=1}^{2}P^{*(i)}_{k1}(\bm{\theta})P^{(j)}_{k1}(\bm{\theta})
				\nonumber\\&
			\times\int\text{d} \omega \phi^*_i(\omega)\phi_j(\omega)\text{e}^{\text{i} \omega(\tau_i-\tau_j)}.\label{eq:overlap_2}
\end{align}
Substituting it back into Eq.~(\ref{eq:overlap_term}) yields
that~$\mathsf{B}_\sigma$, as well as the rate matrix~$R$,
depend on both time delay~$\bm{\tau}$ and polarization~$\bm{\theta}$.

As the permutation matrices $\{\Pi_\sigma\}$ are  unaffected by including polarization parameters,
both rate matrices in Eqs.~(\ref{eq:coincidence_one}) and~(\ref{eq:coincidence_original}) are further reduced by simultaneously diagonalizing class operators of symmetry group~S$_n$~\cite{Chen2002}.
From this reduction procedure,
we obtain a basis transformation $V$ to transform the basis to the basis with permutational symmetry,
which is a linear combination of immanants.

The general formulas of immanant for a matrix~$W$ are
\begin{equation}
  \text{Imm}_\lambda(W):=\sum_{\sigma\in S_n}\chi_\lambda(\sigma)\prod_{i=1}^nW_{i\sigma(i)},
\end{equation}
where~$\lambda$ is a partition of~$n$ and represents an irreducible representation (irrep) of~S$_n$;
$\chi_\lambda$ is the character of~S$_n$ in~$\lambda$ irrep.
Thus, Eq.~(\ref{eq:coincidence_original}) is further written as
\begin{equation}
  \label{eq:coincidence}
  C_{\bm\mu}(\bm{\theta};\bm{\tau})=(V\bm{u})^\dagger \left(VR(\bm{\theta};\bm\tau)V^\dagger\right)(V\bm{u}).
\end{equation}
By writing the coincidence rate in this reduced form,
both the interferometer and the photons are treated in their respective irrep subspaces.
Thus, the calculation is simplified when photons or interferometers are biased with specific permutational symmetries (see Appendix~\ref{app:equivalence}).

\section{Triad Phase Demystified} 
\label{sec:triad_phase_Demystified}
In this section,
we give the specific example for $m=3$ and the scattering network is polarization-insensitive:
\begin{equation}
	\tilde{T}^{(1)}=\tilde{T}^{(2)}=\tilde{T},
\end{equation}
and there is no more than one photon in each spatial mode. 
First we give the triad phase in our framework. 
Each element of the distinguishability matrix of~\cite{Tichy2015PRA,Menssen2017PRL} is our~$\beta$ (see Appendix~\ref{app:equivalence}).

In our framework, the triad phase is defined as the argument of the product of three inner products of two photons' states~\cite{Menssen2017PRL}, 
\begin{equation}\label{eq:triadphase}
	\varphi
		:=\arg\left(\beta_{31}\beta_{12}
			\beta_{23}\right).
\end{equation}
According to Eqs.~(\ref{eq:overlap_term}), (\ref{eq:overlap_1}), and~(\ref{eq:overlap_2})
for the polarization-sensitive case,
\begin{equation}
	\varphi
		=\arg\left(\mathsf{B}(\bm{\tau})_{(123)}\right)\label{eq:sensitive_triad}
\end{equation}
and, for the polarization-insensitive case,
\begin{equation}
	\varphi=\arg\left(\mathsf{B}(\bm{\theta};\bm{\tau})_{(123)}\right)\label{eq:insensitive_traid}.
\end{equation}
In both cases, the triad phase is the argument of the states' overlap of the three-cycle permutation in the rate matrix~$R$.

We also notice that there is a constraint for the triad phase to be an independent variable in the three-photon interference: three photons are identical with the same symmetric density profile around the central frequency~\cite{Menssen2017PRL}.
In that case, 
\begin{equation}
	\operatorname{arg}(\mathsf{B}(\bm{\tau})_{(123)})
		=0.
\end{equation}
Thus, with a symmetric density profile,
for the polarization-sensitive detector,
the triad phase is identically zero.
For the polarization-insensitive detector,
the triad phase is simplified so it is only dependent on the polarization.
Thus, the triad phase inherits the independence from the polarization parameters~$\bm{\theta}$.

Hence, instead of regarding the triad phase as an independent variable to determine multiphoton coincidence,
we see that the polarization parameter itself plays a fundamental role in multiphoton interference. 
We notice that the polarization and the time delays are both tunable parameters to adjust the photons' distinguishability; 
the difference is that the submatrix~$\tilde U$
depends on~$\bm{\theta}$,
but not on $\bm{\tau}$,
so that continuously changing~$\bm{\theta}$
amounts to exploring a parametrized family of $3\times 3$ submatrices $\tilde U(\bm{\theta})$,
which affects the coincidence rates
independently from $\bm{\tau}$.

For the polarization-insensitive case, 
we can mimic the coincidence rate~(\ref{eq:coincidence_original}) using three identically polarized photons of spectra $\phi_1^\prime(\omega)$, $\phi_2^\prime(\omega)$, $\phi_3^\prime(\omega)$ in the original SU(3) interferometer~$\tilde{T}$. 
Notice here that the spectra $\phi^\prime$ is different from $\phi$, 
and should not have an identical symmetric density profile because otherwise there is no argument in the three-cycle permutations.
Then,
\begin{equation}
  \label{eq:equivalence}
  \mathsf{B}^\prime_\sigma({\bm\tau^\prime})=\mathsf{B}_\sigma(\bm{\theta};\bm{\tau}),\, \text{for}\, \forall \sigma\in \text{S}_3.
\end{equation}
The $3!$ complex equations above
correspond to $2\times 3!$ real equations.
In these $2\times 3!$ equations,
the equation where $\sigma=e$ is trivial.

In the remaining equations,
the equations for two-cycle and for three-cycle permutations are not independent.
Thus, the number of independent equations is reduced to four:
two from one complex equation for three-cycle permutation and the rest from the two real equations for two-cycle permutation. 
In addition to the three parameters in~$\bm{\tau}^\prime$, 
another parameter is needed for the four independent equations. 
The triad phase~\cite{Menssen2017PRL} is one such choice.
However, 
here, the photon's spectra are not identical. 
The triad phase~$\varphi$
will be dependent on~$\bm{\tau}^\prime$. 
Thus, we choose one parameter in the photon's spectra, 
for instance, the variance of a Gaussian profile. 
Then these four independent parameters are determined from Eq.~(\ref{eq:equivalence}).

In summary, the triad phase is not a well-defined independent variable in multiphoton interference. The effect of adding a polarization degree of freedom into multiphoton interference is to change the effective SU($2m$) transformation network. 
For the polarization-insensitive detection events,
$\bm{\theta}$ is absorbed into our rate matrix or the two-photon pairwise distinguishability~$\beta$;
thus changing~$\bm{\theta}$ is effectively changing the indistinguishabilities between photons.

\section{Conclusion}
We have described a model for polarized multiphoton interference in a generalized scattering interferometer,
in which the detectors are either polarization sensitive or insensitive.
In our model, polarization is merged into the transformation network by doubling the ports to realize an SU($2m$) transformation.
By changing polarizations,
we consequently change the scattering matrix.

Specifically, for polarization-insensitive detectors, the polarization is merged into our rate matrix, then changing the polarization is equivalently changing the photons' distinguishabilities. 
We have shown that the two-photon pairwise distinguishabilities, which include the information about the independent parameters $\bm{\tau}$ and~$\bm{\theta}$, 
already suffice to describe the multiphoton interference.

As the permutation matrices $\Pi_\sigma$ are not affected by the introduction of the polarization degree of freedom 
(as shown explicitly for the $3\times 3$ case in Appendix~\ref{app:equivalence}), 
block diagonalization of $R(\bm{\theta},\bm{\tau})$ 
yields the coincidence rate in terms of immanants. 
In a generalization of previous results, 
the distinguishability of a photon is not limited to the temporal overlap of pulses but also include the polarization overlap. 
In the specific case of three polarized photons, 
this means that, when two photons are indistinguishable in time and in polarization, 
the rate will contain a sum of moduli squared of the mixed symmetry $\{2,1\}$ immanant and the permanent, 
with no contribution from the determinant of the rate matrix $R(\bm{\theta},\bm{\tau})$. 

As we scale up to~$n$ photons, 
including input and output states which may contain more than one photon, 
the subset of immanants that can appear is determined by the distinguishability of the input photons 
and computed using permutational symmetry methods as per~\cite{Khalid2018PRA}. 
This extends the usefulness of the symmetric group in multiphoton interferometry to situations including polarization.

\section{Acknowledgments}
The work of HdG is supported by NSERC of Canada.  HdG would like to thank Dylan Spivak for clarifying discussions.
BCS acknowledges China's 1000 Talent Plan and NSFC (Grant No.\
11675164) for support.
JW thanks Zheng-Da Li and Hui Wang for valuable discussions.

\appendix
\section{Simplification of coincidence rate of polarization-insensitive detection}
\label{app:polarization_coincidence}
In this appendix, we derive Eq.~(\ref{eq:coincidence_original}) in the main context for the coincidence of polarized photons.
With proposition 1 in~\cite{Khalid2018PRA}, for the measurement in Eq.~(\ref{eq:measure}) we obtain the coincidence rate as
\begin{equation}
C_{\bm\mu}(\bm{\theta};\bm{\tau}):=\sum_{\bm{\eta}\in \mathfrak{S}_{\bm\mu}}C_{\bm\eta}(\bm{\theta};\bm{\tau})
	=\sum_{\bm{\eta}\in \mathfrak{S}_{\bm\mu}}
    	\bm{u}_{\bm\eta}^\dagger(\bm{\theta}) R(\bm\tau)\bm{u}_{\bm\eta}(\bm{\theta}).
\end{equation}
Each entry in $\bm u_{\bm\eta}$ is a function of the elements in the submatrix of~$U$.
From Eqs.~(\ref{eq:UTP}) and~(\ref{eq:explicit_U}),
\begin{equation}
	U_{ij}
  	=\tilde{T}_{\Gamma(i),\Gamma(j)}
    P_{\gamma(i),\gamma(j)}^{(\Gamma(j))},
\end{equation}
is an explicit element of~$U$,
where
\begin{equation}
	\gamma(i)
    	=\begin{cases}
  1 \, &i\leq m,\\
  2 \, &\text{otherwise}.
\end{cases}
\end{equation}
indicates the polarization. 
On the other hand,
\begin{equation}
	\Gamma(i)=i+m-m\gamma(i)
\end{equation}
indicates which photon the state belongs to.
Thus, when $v_i\leq m$ for all injected photons, and $\bm{\eta}\in \mathfrak{S}_{\bm\mu}$, we obtain
\begin{equation}
	\gamma(v_i)=1,\;
	\Gamma(v_i)=v_i,
	\Gamma(\bm{\eta})=\bm{\mu}.
\end{equation}
Thus,  each element in $\bm{u_\mu}$ is
\begin{equation}
(\bm{u}_{\bm\eta})_k:=\prod_{i=1}^n U_{\bar{\eta}_i^{k},v_i}(\bm{\theta})=\prod_{i=1}^n \tilde{T}_{\bar{\mu}_i^k, v_i}P^{(v_i)}_{\gamma(\bar{\eta}_i^k),1}(\bm{\theta}).
\end{equation}
Then we write the polarization part out from $\bm{u}_{\bm\eta}$ as a product of a diagonal matrix $K_{\bm\eta}$,
with diagonal elements
\begin{equation}
  \left[K_{\bm\eta}(\bm{\theta})\right]_{kk}=\prod_{i=1}^nP^{(v_i)}_{\gamma(\bar{\eta}_i^k),1}(\bm{\theta}),\,\text{for}\, k=1,2,\dots,N,
\end{equation}
and~$\bm{u}$ with elements
\begin{equation}
	u_k
    =\prod_{i=1}^n
    	\tilde{T}_{\bar{\mu}_i^k, v_i}.\label{eq:A6}
\end{equation}
We first sum all output configurations with the rate matrix according to 
\begin{align}
  \sum_{\bm\eta\in\mathfrak{S}_\mu}
  	K_{\bm\eta}^\dagger(\bm{\theta}) &R(\bm{\tau})K_{\bm\eta}(\bm{\theta})
  			\nonumber\\
  	=&\sum_{\sigma\in S_n}\mathsf{B}_\sigma(\bm{\tau})
		\sum_{\bm\eta\in\mathfrak{S}_\mu}K_{\bm\eta}^\dagger(\bm{\theta})\Pi_\sigma K_{\bm\eta}(\bm{\theta})\label{eq:A7}.
\end{align}
The last summation is simplified to
\begin{align}\notag
&\left(\sum_{\bm{\eta}\in\mathfrak{S}_\mu}K_{\bm\eta}^\dagger(\bm{\theta})\Pi_\sigma K_{\bm\eta}(\bm{\theta})\right)_{pq}\\\notag
	=&\sum_{\bm{\eta}\in\mathfrak{S}_\mu}\left[K^*_{\bm\eta}(\bm{\theta})\right]_{pp} (\Pi_\sigma)_{pq} \left[K_{\bm\eta}(\bm{\theta})\right]_{qq}\\\notag
	=&\sum_{\bm{\eta}\in\mathfrak{S}_\mu}\prod_{i=1}^nP^{*(v_i)}_{\gamma(\bar{\eta}_i^p),1}(\bm{\theta})\delta(\sigma\bar{\bm{\eta}}^{p},\bar{\bm{\eta}}^{q})\prod_{j=1}^nP^{(v_j)}_{\gamma(\bar{\eta}_j^q),1}(\bm{\theta})\\
	=&\begin{cases}
\sum_{\bm{\eta}\in\mathfrak{S}_\mu}\prod_{i=1}^nP^{*(v_{\sigma(i)})}_{\gamma(\bar{\eta}_i^p),1}(\bm{\theta})P^{( v_i)}_{\gamma(\bar{\eta}_i^p),1}(\bm{\theta}), &\sigma\bar{\bm{\eta}}^{p}
	=\bar{\bm{\eta}}^{q},\\
0,&\text{otherwise}.
\end{cases}\label{eq:A10}
\end{align}
And,
\begin{align}\notag
&\sum_{\bm\eta\in\mathfrak{S}_\mu}\prod_{i=1}^nP^{*(v_{\sigma(i)})}_{\gamma(\bar{\eta}_i^p),1}(\bm{\theta})P^{(v_i)}_{\gamma(\bar{\eta}_i^{p}),1}(\bm{\theta})\\
	&=\prod_{i=1}^{n}\left[P_{11}^{*(v_{\sigma(i)})}(\bm{\theta})P_{11}^{(v_i)}(\bm{\theta})+P_{21}^{*(v_{\sigma(i)})}(\bm{\theta})P_{21}^{(v_i)}(\bm{\theta})\right]\label{eq:A11}.
\end{align}
Thus, together with Eqs.~(\ref{eq:A6}),~(\ref{eq:A7}),~(\ref{eq:A10}),
and~(\ref{eq:A11}), 
we obtain the coincidence rate,
\begin{equation}
C_{\bm\mu}(\bm{\tau})=\bm{u}^\dagger R(\bm{\theta};\bm{\tau})\bm{u},
\end{equation}
with
\begin{equation}
R(\bm{\theta};\bm{\tau}):=\sum_{\sigma\in S_n}\mathsf{B}_\sigma(\bm{\theta};\bm{\tau})\Pi_\sigma,
\end{equation}
and
\begin{align}
\mathsf{B}_\sigma(\bm{\theta};\bm{\tau})&=\prod_{i=1}^{n}\sum_{k=1}^2P_{k1}^{*(v_{\sigma(i)})}(\bm{\theta})P_{k1}^{(v_i)}(\bm{\theta})\mathsf{B}_\sigma(\bm{\tau})\\
&=\prod_{i=1}^n\beta_{v_{\sigma(i),v_i}}(\bm{\theta};\bm{\tau}),
\end{align}
with~$\beta_{ij}$ defined in Eq.~(\ref{eq:overlap_2}).

\begin{widetext}
\section{Equivalence between permanent-based formalism and immanant-based formalism}
\label{app:equivalence}
For simplification, 
we verify the equivalence between the immanant-based formalism before the block diagonalization 
and the permanent-based formalism before tracing over one permutation index
for the case when each mode is occupied by no more than one photon. 
The general case is verified afterwards. 
Specifically, we verify the equivalence between
\begin{equation}
	C(\bm{\theta};\bm{\tau})=\bm{u}^\dagger R(\bm{\theta};\bm{\tau}) \bm{u}
\end{equation}
and Eq.~(19) in~\cite{Tichy2015PRA}.

We inject~$n$ photons into an optical interferometer $U\in \text{SU}(m)$ with input configuration~$\bm{v}$.
First we notice that Eqs.~(4) and (6) in~\cite{Tichy2015PRA}
use a different definition of scattering matrix~$U$.
To demonstrate equivalence,
we employ the definition in~\cite{Tichy2015PRA} as the first index represents the input and the second index represents the output.
Thus,
\begin{equation}
	a^\dagger_i \mapsto \sum_{j=1}^{m}U_{ij}a^\dagger_j,
\end{equation}
We also have
\begin{equation}
	u_k=\prod_{i=1}^n U_{v_i \mu_{\sigma^k(i)}},
\end{equation}
where $\sigma^k$ is the $k$th element in symmetric group~S$_n$, and 
\begin{align}
	R_{ij}(\bm{\theta};\bm{\tau})&=\sum_{\sigma\in \text{S}_n}\mathsf{B}_\sigma(\bm{\theta};\bm{\tau})(\Pi_\sigma)_{ij}\\
&=\sum_{\sigma\in \text{S}_n}\mathsf{B}_\sigma(\bm{\theta};\bm{\tau})\delta(\sigma \sigma^i,\sigma^j)=\mathsf{B}_{\sigma^j(\sigma^i)^{-1}}(\bm{\theta};\bm{\tau}).
\end{align}
We substitute them into the coincidence rate to obtain
\begin{align}\notag
	&C(\bm{\theta};\bm{\tau})
=\sum_{i,j}u_i^*R_{ij}(\bm{\theta};\bm{\tau})u_j
=\sum_{i,j}\prod_{k=1}^n U^*_{v_k \mu_{\sigma^i(k)}}\mathsf{B}_{\sigma^j(\sigma^i)^{-1}}(\bm{\theta};\bm{\tau})\prod_{l=1}^n U_{v_l \mu_{\sigma^j(l)}}\\
	=&\sum_{i,j}\prod_{k=1}^n U^*_{v_k \mu_{\sigma^i(k)}}U_{v_k \mu_{\sigma^j(k)}}\prod_{l}\beta_{v_{\sigma^j(\sigma^i)^{-1}(l)},v_l}(\bm{\theta};\bm{\tau})
	=\sum_{i,j}\prod_{k=1}^n U^*_{v_k \mu_{\sigma^i(k)}}U_{v_k \mu_{\sigma^j(k)}}\beta_{v_{(\sigma^i)^{-1}(k)},v_{(\sigma^j)^{-1}(k)}}(\bm{\theta};\bm{\tau})\\
=&\sum_{i,j}\prod_{k=1}^n U^*_{v_{(\sigma^i)^{-1}(k)} \mu_{(k)}}U_{v_{(\sigma^j)^{-1}(k)} \mu_{(k)}}\beta_{v_{(\sigma^i)^{-1}(k)},v_{(\sigma^j)^{-1}(k)}}(\bm{\theta};\bm{\tau}).
\end{align}
If we define our new permutation operator~$\rho$ as $\rho=\sigma^{-1}$, then
\begin{equation}
C(\bm{\theta};\bm{\tau})=\sum_{i,j}\prod_{k=1}^n U^*_{v_{\rho^i(k)}\eta(k)}U_{v_{\rho^j(k)}\eta(k)}\beta_{v_{\rho^i(k)},v_{\rho^j(k)}}(\bm{\theta};\bm{\tau})\label{eq:same_Tichy}
\end{equation}
agrees with Eq.~(19) in~\cite{Tichy2015PRA}
where we identify
\begin{equation}
	\beta:=\mathcal{S}.
\end{equation}

Let us make this verification explicit with $m=n=3$.
We start with the formalism in~\cite{Tichy2015PRA} or, equivalently, Eq.~(\ref{eq:same_Tichy}).
Then
\begin{equation}
	\bm{v}=(1,2,3),\;
	\bm{\eta}=(1,2,3).
\end{equation}
The summation of Eq~(\ref{eq:same_Tichy}) contains~36 terms,
which is ultimately expressed compactly as 
\begin{align}
	C=&\begin{pmatrix}
U_{11}U_{22}U_{33}\\
U_{11}U_{23}U_{32}\\
U_{12}U_{21}U_{33}\\
U_{12}U_{23}U_{31}\\
U_{13}U_{21}U_{32}\\
U_{13}U_{22}U_{31}
\end{pmatrix}^\dagger\begin{pmatrix}
\beta_{11}\beta_{22}\beta_{33}&\beta_{11}\beta_{23}\beta_{32}&\beta_{12}\beta_{21}\beta_{33}&\beta_{13}\beta_{21}\beta_{32}&\beta_{12}\beta_{23}\beta_{31}&\beta_{13}\beta_{22}\beta_{31}\\
\beta_{11}\beta_{32}\beta_{23}&\beta_{11}\beta_{33}\beta_{22}&\beta_{12}\beta_{31}\beta_{23}&\beta_{13}\beta_{31}\beta_{22}&\beta_{12}\beta_{33}\beta_{21}&\beta_{13}\beta_{32}\beta_{21}\\
\beta_{21}\beta_{12}\beta_{33}&\beta_{21}\beta_{13}\beta_{32}&\beta_{22}\beta_{11}\beta_{33}&\beta_{23}\beta_{11}\beta_{32}&\beta_{22}\beta_{13}\beta_{31}&\beta_{23}\beta_{12}\beta_{31}\\
\beta_{31}\beta_{12}\beta_{23}&\beta_{31}\beta_{13}\beta_{22}&\beta_{32}\beta_{11}\beta_{23}&\beta_{33}\beta_{11}\beta_{22}&\beta_{32}\beta_{13}\beta_{21}&\beta_{33}\beta_{12}\beta_{21}\\
\beta_{21}\beta_{32}\beta_{13}&\beta_{21}\beta_{33}\beta_{12}&\beta_{22}\beta_{31}\beta_{13}&\beta_{23}\beta_{31}\beta_{12}&\beta_{22}\beta_{33}\beta_{11}&\beta_{23}\beta_{32}\beta_{11}\\
\beta_{31}\beta_{22}\beta_{13}&\beta_{31}\beta_{23}\beta_{12}&\beta_{32}\beta_{21}\beta_{13}&\beta_{33}\beta_{21}\beta_{12}&\beta_{32}\beta_{23}\beta_{11}&\beta_{33}\beta_{22}\beta_{11}
\end{pmatrix}\begin{pmatrix}
U_{11}U_{22}U_{33}\\
U_{11}U_{23}U_{32}\\
U_{12}U_{21}U_{33}\\
U_{12}U_{23}U_{31}\\
U_{13}U_{21}U_{32}\\
U_{13}U_{22}U_{31}
\end{pmatrix},
\end{align}
ignoring the indices~$\bm{\theta}$ and $\bm{\tau}$ for clarity.
Introducing the permutation matrices

\begin{align}
	\Pi_{(23)}
		=\begin{pmatrix}
0 & 1 & 0 & 0 & 0 & 0\\
1 & 0 & 0 & 0 & 0 & 0\\
0 & 0 & 0 & 1 & 0 & 0\\
0 & 0 & 1 & 0 & 0 & 0\\
0 & 0 & 0 & 0 & 0 & 1\\
0 & 0 & 0 & 0 & 1 & 0\\
\end{pmatrix},
\Pi_{(12)}=\begin{pmatrix}
0 & 0 & 1 & 0 & 0 & 0\\
0 & 0 & 0 & 0 & 1 & 0\\
1 & 0 & 0 & 0 & 0 & 0\\
0 & 0 & 0 & 0 & 0 & 1\\
0 & 1 & 0 & 0 & 0 & 0\\
0 & 0 & 0 & 1 & 0 & 0\\
\end{pmatrix},\label{eq:P12matrix}
\Pi_{(123)}
	=\begin{pmatrix}
0 & 0 & 0 & 1 & 0 & 0\\
0 & 0 & 0 & 0 & 0 & 1\\
0 & 1 & 0 & 0 & 0 & 0\\
0 & 0 & 0 & 0 & 1 & 0\\
1 & 0 & 0 & 0 & 0 & 0\\
0 & 0 & 1 & 0 & 0 & 0\\
\end{pmatrix},
\end{align}
\begin{align}
\Pi_{(132)}
	=\begin{pmatrix}
0 & 0 & 0 & 0 & 1 & 0\\
0 & 0 & 1 & 0 & 0 & 0\\
0 & 0 & 0 & 0 & 0 & 1\\
1 & 0 & 0 & 0 & 0 & 0\\
0 & 0 & 0 & 1 & 0 & 0\\
0 & 1 & 0 & 0 & 0 & 0\\
\end{pmatrix},
	\Pi_{(13)}
		=\begin{pmatrix}
0 & 0 & 0 & 0 & 0 & 1\\
0 & 0 & 0 & 1 & 0 & 0\\
0 & 0 & 0 & 0 & 1 & 0\\
0 & 1 & 0 & 0 & 0 & 0\\
0 & 0 & 1 & 0 & 0 & 0\\
1 & 0 & 0 & 0 & 0 & 0\\
\end{pmatrix},
\end{align}
the rate is
\begin{align}
	C&=\beta_{11}\beta_{22}\beta_{33}\openone+\beta_{11}\beta_{23}\beta_{32}\Pi_{(23)}+\beta_{12}\beta_{21}\beta_{33}
\Pi_{(12)}\nonumber+\beta_{13}\beta_{21}\beta_{32}\Pi_{(123)}+\beta_{12}\beta_{23}\beta_{31}\Pi_{(132)}
+\beta_{13}\beta_{22}\beta_{31}
\Pi_{(13)}\label{eq:rep}.
\end{align}
Block diagonalization of the permutation matrices, which span in this specific case the regular
representation of S$_3$,
produces linear combinations of $U_{1i}U_{2j}U_{3k}$ related to immanants.
Consequently,
block diagonalization provides additional information not contained in Eq.~(\ref{eq:same_Tichy}) related to the
permutation structure of the partially distinguishable particles of the system.
Importantly,
block diagonalization does not depend on the nature or the origin of the~$\beta_{ij}$ coefficients.

To be explicit,
consider the matrix
\begin{equation}
V=
\frac{1}{\sqrt{6}}
\begin{pmatrix}
 1 & 1 & 1 & 1 & 1 & 1 \\
 1 & -1 & -1 & 1 & 1 & -1 \\
 \sqrt{\frac{3}{2}} & \sqrt{\frac{3}{2}} & -\sqrt{\frac{3}{2}} & 0 & -\sqrt{\frac{3}{2}} & 0 \\
 -\frac{1}{\sqrt{2}} & -\frac{1}{\sqrt{2}} & -\frac{1}{\sqrt{2}} & \sqrt{2} & -\frac{1}{\sqrt{2}}
   & \sqrt{2} \\
 \sqrt{\frac{3}{2}} & -\sqrt{\frac{3}{2}} & \sqrt{\frac{3}{2}} & 0 & -\sqrt{\frac{3}{2}} & 0 \\
 \frac{1}{\sqrt{2}} & -\frac{1}{\sqrt{2}} & -\frac{1}{\sqrt{2}} & -\sqrt{2} & \frac{1}{\sqrt{2}}
   & \sqrt{2} \\
\end{pmatrix} \label{eq:Vblockdiagonal}
\end{equation}
This matrix
(obviously independent of the details of the interferometer)
transforms the rate matrix $R(\bm{\theta};\bm{\tau})$,
containing products of $\beta_{ij}$'s as entries,
to block-diagonal form.
Using the notation $\beta_{ij}=r_{ij}\text{e}^{\text{i}\phi_{ij}}$ as per~\cite{Menssen2017PRL} yields
\begin{equation}
V R(\bm{\theta};\bm{\tau}) V^{-1}=
\begin{pmatrix}
{\cal P}&0&0&0&0&0\\
0&{\cal D}&0&0&0&0\\
0&0&{\cal A}&{\cal B}&0&0\\
0&0&{\cal B}^*&{\cal C}&0&0\\
0&0&0&0&{\cal C}&{\cal B}^*\\
0&0&0&0&{\cal B}&{\cal A}\\
\end{pmatrix}
\end{equation}
where
\begin{align}
{\cal A}&=1+\frac{r_{13}^2}{2}+\frac{r_{23}^2}{2}-r_{12}^2-r_{12} r_{13} r_{23} \cos \varphi\\
{\cal B}&= \frac{\sqrt{3}}{2}  \left(r_{13}^2-r_{23}^2-2 \text{i} r_{12} r_{13} r_{23} \sin \varphi\right)\\
{\cal C}&=1-\frac{r_{13}^2}{2}-\frac{r_{23}^2}{2}+r_{12}^2-r_{12} r_{13} r_{23} \cos \varphi\\
{\cal P}&=1+r_{13}^2+r_{23}^2+r_{12}^2+2 r_{12} r_{13} r_{23} \cos \varphi\\
{\cal D}&= 1-r_{13}^2-r_{23}^2-r_{12}^2+2 r_{12} r_{13} r_{23} \cos \varphi
\end{align}
and $\varphi=\phi_{31}+\phi_{12}+\phi_{23}$ is the triad phase consistent with the definition in Eq.~(\ref{eq:triadphase})
and where
\begin{equation}
	\phi_{31}=-\phi_{13}
\end{equation}
has been used.
The transformation $V$ induces a change from:
\begin{equation}
\bm{u}=\begin{pmatrix}
U_{11}U_{22}U_{33}\\
U_{11}U_{23}U_{32}\\
U_{12}U_{21}U_{33}\\
U_{12}U_{23}U_{31}\\
U_{13}U_{21}U_{32}\\
U_{13}U_{22}U_{31}
\end{pmatrix}\to
\tilde{\bm{u}}=V\bm{u}
\end{equation}
where $\tilde{u}_{1}$ is the permanent of $U(\bm{\theta};\bm{\tau})$,
$\tilde{u}_{2}$ is the determinant of $U(\bm{\theta};\bm{\tau})$
and the remaining components are linear combination of immanants for 
the irrep $\set{2,1}$ of S$_{3}$ of the matrix $U(\bm{\theta};\bm{\tau})$ with some columns permuted~\cite{Tillmann2015PRX, deGuise2014PRA}.

The transformation $V$ also happens to diagonalize the permutation matrix $\Pi_{(12)}$ of Eq.~(\ref{eq:P12matrix}):
\begin{align}
V \Pi_{12} V^{-1}=\left(
\begin{array}{cccccc}
 1 & 0 & 0 & 0 & 0 & 0 \\
 0 & -1 & 0 & 0 & 0 & 0 \\
 0 & 0 & -1 & 0 & 0 & 0 \\
 0 & 0 & 0 & 1 & 0 & 0 \\
 0 & 0 & 0 & 0 & 1 & 0 \\
 0 & 0 & 0 & 0 & 0 & -1 \\
\end{array}
\right)
\label{eq:blockdiagonalP12}
\end{align}
which singles out particle 1 and particle 2 for additional simple analysis.
Indeed, if we make these two particles indistinguishable,
so that,
\begin{equation} 
r_{12} \to 1, \quad r_{13}\to r_{23},\quad \phi_{12}\to 0,\quad \phi_{31}\to -\phi_{23}
\label{eq:2samecondition}
\end{equation}
then the matrix $VR(\bm{\theta};\bm{\tau})V^{-1}$ collapses to a particular simple form,
where
\begin{align}
V &\,R(\bm{\theta};\bm{\tau}) V^{-1}\to
\begin{pmatrix}
2+4r_{23}^2 & 0 & 0 & 0 & 0 & 0 \\
0 & 0 & 0 & 0 & 0 & 0 \\
0 & 0 & 0 & 0 & 0 & 0 \\
0 & 0 & 0 & 2-2 r_{23}^2 & 0 & 0 \\
0 & 0 & 0 & 0 & 2-2r_{23}^2 & 0 \\
0 & 0 & 0 & 0 & 0 & 0
\end{pmatrix}  
\label{eq:Rblockdiagonal}
\end{align}
showing that consistent with previous analysis,
only the permanent of the original scattering matrix $U(\bm{\theta};\bm{\tau})$
plus some combinations of immanants for the irrep $\set{2,1}$ will survive.

Finally, we note that
if bosons $1$ and $2$ are indistinguishable in the input state $\vert \Psi_{\text{in}}\rangle$, 
this state will be an eigenstate of $\Pi_{12}$ with eigenvalue $+1$.  
It is therefore completely clear why only the entries of 
$VR(\bm{\theta};\bm{\tau}) V^{-1}$ in Eq.~(\ref{eq:Rblockdiagonal}) that remain after the substitution of Eq.~(\ref{eq:2samecondition}) are those corresponding to the eigenvalues $+1$ in Eq.~(\ref{eq:blockdiagonalP12}).
\end{widetext}
\bibliography{TriadPhase}
\end{document}